\documentclass[aps,prb,twocolumn,amsmath,amssymb,floatfix]{revtex4}
\usepackage{graphicx}
\usepackage{dcolumn}
\usepackage{bm}
\usepackage{amsfonts}
\usepackage{amsmath}
\usepackage{ulem}
\usepackage{color}
\usepackage{bigints}
\usepackage{tensor}

\newcommand{\ie}{{\it i.e.\ }}

\newcommand{\nphys}{{Nat. Phys. }}
\newcommand{\re}{{\Re}{\rm e}}

\begin{document}

\title{Majorana qubit decoherence by quasiparticle poisoning} 

\author{Diego Rainis}
\author{Daniel Loss}
\affiliation{Department of Physics, University of Basel, Klingelbergstrasse 82, 4056 Basel, Switzerland}

\begin{abstract}
We consider the problem of quasiparticle poisoning in a nanowire-based realization of a Majorana qubit, where a spin-orbit-coupled semiconducting wire is placed on top of a (bulk) superconductor. By making use of recent experimental data exhibiting evidence of a low-temperature residual non-equilibrium quasiparticle population in superconductors, we show by means of analytical and numerical calculations that the dephasing time due to the tunneling of quasiparticles into the nanowire may be problematically short to allow for qubit manipulation. 
\end{abstract}

\maketitle
Devising a physical system where to experimentally observe for the first time the presence of Majorana fermions has become in the last years a serious and stirring challenge in the condensed-matter physics community.
Apart from the importance {\it per se} of observing the signature a Majorana fermion, the strong hope is to realize a Majorana-based qubit, which would offer an intrinsically improved protection against decoherence due to the peculiar delocalized structure of the Majorana state itself. 
For example, in the setup we consider,\cite{KitaevPU2001,Sato,proposals,AliceaPRB2010} a topological superconducting state (TSC) in a semiconducting nanowire is created, with a mid-gap mode $d^\dagger_{\rm end}$ at energy $\varepsilon_{\rm end}\simeq 0$, whose wavefunction is strongly localized at the two ends of the nanowire. 
Such topological state can be induced through the combined effect of $s-$wave pairing, spin-orbit coupling and magnetic field~\cite{Sato,proposals,AliceaPRB2010}. The superconducting pairing is inherited most typically via proximity effect from a bulk superconductor placed below the wire (around the wire in some proposals).
Even without restricting ourselves to this specific setup, superconductivity is a key ingredient needed in essentially all the proposals to produce observable Majorana excitations in condensed-matter systems.
The considerations we make in this paper are thus qualitatively valid and relevant for a wide range of configurations, while the quantitative results are specific to the proximized nanowire setup.

The zero-energy many-body excitation $d_{\rm end}$ in the TSC can be exploited to store information in an ideally dephasing-free qubit. Defining the $|0\rangle$ state as the many-body state where the $d_{\rm end}$ is empty, and correspondingly $|1\rangle\equiv d_{\rm end}^\dagger|0\rangle$, the subspace spanned by $|0\rangle$ and $|1\rangle$ is a degenerate ground-state subspace, which offers intrinsic protection against dephasing. 
However, coherent states of the type $\left(|0\rangle + {\rm e}^{i\phi}|1\rangle\right)$ cannot be prepared, because there is no physical coupling which could create such superposition. 
Strictly speaking, then, a system where the states $|0\rangle$ and $|1\rangle$ differ by fermion parity (occupation of a single BCS-like mode) cannot be used as a qubit.
To obtain a proper quantum bit, one needs at least two of these zero-energy states, that is, four Majorana fermions. In such case, 
there are four degenerate states:
\begin{align}
\begin{array}{rl}
|00\rangle&\equiv |0\rangle_{\rm end,1}\otimes|0\rangle_{\rm end,2}\;, \nonumber \\
|10\rangle&\equiv d_{\rm end,1}^\dagger|00\rangle\;, \nonumber  \\ 
|01\rangle&\equiv d_{\rm end,2}^\dagger|00\rangle\;, \nonumber  \\
|11\rangle&\equiv d_{\rm end,1}^\dagger d_{\rm end,2}^\dagger|00\rangle\;. 
\end{array}
\end{align}
The states $|00\rangle$ and $|11\rangle$ (possible choice for the qubit) 
share now the same fermion parity, and if we choose them as qubit computational states, coherent superpositions are possible thanks to the superconducting pairing which induces fluctuations in the number of electrons, { in jumps of two}, due to the hopping in and out of Cooper pairs.

If the superconductor can only exchange Cooper pairs and not single, unpaired electrons with the wire, then the fermion parity (\ie the number of electrons modulo two) is conserved. This is at the base of the protection these systems benefit~\cite{FuKanePRL2008,AkhmerovPRB2010}. If instead single electrons could enter the nanowire, then the system would be driven out of the topological subspace $\left\{|00\rangle,|11\rangle \right\}$, populating $|10\rangle$, $|01\rangle$ or some higher-energy states.
In the case of a single TSC segment, with only one zero-energy mode, the presence of unpaired electrons would instead cause $\sigma_x$ errors, causing transitions $|0\rangle\rightarrow |1\rangle$ and viceversa.

There have already been a couple of works~\cite{Chamon2011PRB,Budich2012PRB} pointing out that Majorana-based qubits are prone to standard decoherence mechanisms when one allows for single-electron tunneling from a generic external (noisy) environment.
The specific phenomenon of the possible disturbing presence of unpaired electrons in the superconductor, dubbed ``quasiparticle poisoning'', constitutes a similar but more subtle problem, since the tunneling electrons are coming from the superconductor itself, which is an indispensable ingredient in these type of proposals.
The issue has already been taken into consideration in the Majorana-fermion community~\cite{vanHeckPRB2011,FuKanePRB2009,HasslerNJP2011} but, to the best of our knowledge, no explicit estimation has been made for a specific setup. 
Rather, some estimates for quasiparticle tunneling rate from the superconductor into the wire have been borrowed from some recent experiments dealing with equilibrium superconductivity~\cite{deVisserPRL2011}. 
In such experiments, like in several other investigations before~\cite{AumentadoPRL2004,FergusonPRL2006,ShawPRB2008,EstevePRL2011},
low-temperature quasiparticle poisoning has been observed, and evidence has been provided that the quasiparticle density does not drop to zero at the smallest temperatures, as predicted by theory, but rather saturates to a constant value. 
A clear and widely accepted explanation for this phenomenon is still lacking, but the observation of excess quasiparticles has been repeatedly confirmed.
For instance, a very recent experimental investigation on quasiparticle kinetics inducing energy relaxation in a transmon qubit has measured poisoning times below $\mu$s~\cite{SchoelkopfArXiv2011}.

In an ideal superconductor at zero temperature all the electrons are forming Cooper pairs and out of the condensate no quasiparticles exist~\cite{tinkham}. 
At finite but small temperature, 
the average density of quasiparticles in a superconductor {\it at equilibrium} follows the activation behavior ($k_{\rm B}=1$)
\begin{equation}
	n_{\rm qp}^{\rm eq} = 2\nu_{\rm S}^{\rm n}\sqrt{2\pi T \Delta_{\rm S}}\exp(- \Delta_{\rm S}/T)\;,
	\label{eq:Nqp}
\end{equation}
valid at $T <  \Delta_{\rm S}$, with $\nu_{\rm S}^{\rm n}$ the normal-state, single-spin density of states at the Fermi level, $T$ the temperature and $ \Delta_{\rm S}$ the energy gap of the superconductor.
The corresponding average quasiparticle recombination time (lifetime) is evaluated as~\cite{KaplanPRB1976}:
\begin{equation}
	\tau_r=
	\frac{\tau_0}{\sqrt{\pi}}\left(\frac{T_{\rm c}}{2  \Delta_{\rm S}}\right)^{\frac{5}{2}}\sqrt{\frac{T_{\rm c}}{T}}~{\rm e}^{{  \Delta_{\rm S}}/{T}}
	\propto \frac{1}{n_{\rm qp}}\;,
	\label{eq:taur}
\end{equation}
where $T_{\rm c}$ is the critical temperature of the superconductor and $\tau_0$ a material dependent, characteristic electron-phonon interaction time. Eqs.~(\ref{eq:Nqp}) and~(\ref{eq:taur}) predict a very low quasiparticle density and correspondingly a very long quasiparticle lifetime at temperatures $T\ll T_{\rm c}$. 
As anticipated above, experimental data show agreement to the exponential behavior Eq.~(\ref{eq:taur}) only for not-too-low temperatures.
For instance, the work Ref.~[\onlinecite{deVisserPRL2011}] obtains a zero-temperature saturation $\tau_{\rm r}\sim 2$ ms in the quasiparticle lifetime for temperatures below 160 mK, and a corresponding saturating quasiparticle density $n_{\rm qp}\sim 25-55~\mu{\rm m}^{-3}$. 
These large quaisparticles lifetimes of the order of ms have been sometimes used in the literature as an estimate for the tunneling rate of quasiparticles into the TSC nanowire, {\it i.e.} an estimate for the qubit lifetime, leading to the conclusion that quasiparticle poisoning is not a serious issue.

Such experiments~\cite{deVisserPRL2011}, however, analyze the quasiparticle lifetime $\tau_{\rm r}$ in an isolated superconducting system (two-dimensional in the case of Ref.~[\onlinecite{deVisserPRL2011}]), and not in a hybrid structure where two subsystems are put into contact, and a subsystem can be poisoned by the other one. 

Here we would like to stress that the rate of tunneling into the qubit subsystem (the nanowire), and thus the average dephasing time of our Majorana qubit, is {\it not} given by the quasiparticles lifetime as measured in an isolated superconductor. Indeed, for example in the case of our SC/TSC junction, quasiparticles in the superconducting reservoir can well tunnel into the TSC wire and tunnel out again, many times before a recombination event may take place after the typical time $\tau_{\rm r}$. Since even a single detour of the quasiparticle into the qubit part of the system could destroy the coherence of the qubit itself, it is important to calculate or measure explicitly the tunneling rate of quasiparticles $\Gamma_{\rm qp}$ into the nanowire. 
To this end, we adopt a well-established formalism, already employed in earlier calculations for $\Gamma_{\rm qp}$ which were performed for the case of Josephson qubits, adapting them to the present case of a SC interfaced to a TSC nanowire. We demonstrate that, depending on the parameters, the tunneling rate of quasiparticles can vary in the range $0.1-100$ MHz, imposing therefore a much more serious constraint on the feasibility of error-free qubit manipulation. This becomes particularly clear in the last section of this paper, where we produce quantitative estimates for the poisoning rate in typical configurations, and we compare them with the time scales required for adiabatic computations.

\section{Calculations}
The system we consider is constituted by a bulk $s-$wave superconductor in tunnel-contact with a TSC nanowire. Gap magnitudes $  \Delta_{\rm S}$ and $\Delta_{\rm T}$ in the SC and in the TSC need not be the same. Rather, the topological gap $\Delta_{\rm T}$ is always smaller than $  \Delta_{\rm S}$, for two reasons: (i) the proximity-induced gap is in general smaller than the parental gap $ \Delta_{\rm S}$, depending on the transparency of the interface SC/TSC~\cite{proximity}; (ii) the topological $p-$wave gap $\Delta_{\rm T}$ is only a fraction of the induced $s-$wave amplitude, and strongly depends on the values of magnetic field, spin-orbit coupling~\cite{AliceaPRB2010}, and even on the electron-electron interaction~\cite{interactions}. 
\\
The bulk superconductor is described by a standard BCS Hamiltonian $H_{\rm SC}$, whose eigen-excitations are bogolons $\beta_{k\sigma}$ with energy $E_{\rm S}(k)=\sqrt{\xi^2_k+\Delta^2_{\rm S}}=\sqrt{\left( \hbar^2 k^2/2m-\mu \right)^2+\Delta^2_{\rm S}}$:
\begin{equation}
H_{\rm SC}=\sum_{k\sigma} E_{\rm S}(k) \beta^\dagger_{k\sigma} \beta_{k\sigma} \;.
\end{equation}
As mentioned above, the nanowire can be driven into a topological superconducting phase by means of the combined effect of spin-orbit coupling, Zeeman interaction and superconducting pairing~\cite{AliceaPRB2010}.
The topological phase is reached by tuning the chemical potential within the gap opened by the Zeeman interaction between the two chirality bands, and singling out in this way a single pair of Fermi points. In terms of this single effective degree of freedom, the original $s-$wave superconducting interaction becomes of $p-$type. The low-energy effective Hamiltonian for such spinless $p-$wave superconductor then reads~\cite{AliceaPRB2010,AliceaNPHYS2011}:
\begin{equation}
H_{\rm NW}=\sum_{k} \left[ \varepsilon_k d^\dagger_{k} d_{k} + {\rm sgn}(k)\left( \Delta_{\rm T}  d^\dagger_{k}d^\dagger_{-k} +\Delta_{\rm T}^*  d_{-k}d_{k}   \right) \right]\;,
\end{equation}
where the $d_k$'s describe the lower-band electrons originating from the combined effect of spin-orbit and Zeeman interaction, with dispersion $ \varepsilon_k$.
After diagonalization, the low-energy Hamiltonian of the nanowire is also written in terms of bogolons $\eta_k$,
\begin{equation}
H_{\rm NW}\rightarrow H_{\rm TSC}=\sum_{k}  E_{\rm T}(k) \eta^\dagger_{k} \eta_{k} \;,
\end{equation}
with dispersion $E_{\rm T}(k)=\sqrt{\varepsilon_k^2+\Delta^2_{\rm T}}$.
Finally, the two subsystems are coupled by tunneling, described by the Hamiltonian $H_{\rm T}$,
\begin{align}\label{HT}
H_{\rm T}&=\sum_{k,p,\sigma} \left( t_{kp}^{(0)} c^\dagger_{k,\sigma} a_{p,\sigma} +  t_{kp}^{(0)*} a^\dagger_{p,\sigma} c_{k,\sigma}\right)\\ \label{HT2}
&=\sum_{k,p,\sigma} \left( t_{kp\sigma} d^\dagger_{k} a_{p,\sigma} + t_{kp\sigma}^* a^\dagger_{p,\sigma} d_{k}\right)\;.
\end{align}
The operators $a_{p,\sigma}$ annihilate an electron in the state $|p,\sigma\rangle$ in the SC reservoir, while the $c_{k,\sigma}$'s are bare-electron operators in the nanowire.
Switching to the diagonal basis of Zeeman and Rashba in the nanowire leads to the final expression (\ref{HT2}) written in terms of
the effective spinless lower-band electron operators $d_{k}$ introduced above.
Here the tunneling amplitudes $t_{kp\sigma}$ differ from the bare-electron tunneling amplitudes $t_{kp}^{(0)}$, since they describe the hopping into the effective spinless modes $d_k$, and incorporate the spin-dependent factors which describe the mixing of degrees of freedom due to spin-orbit and Zeeman interaction.

\section{Estimation without environmental $P(E)$ theory}
We first start with the case where the only relevant degrees of freedom are those related to the electronic quasiparticle tunneling through the SC/TSC junction. In a more refined theory the event of a quasiparticle tunneling through the junction is influenced by the charge dynamics in the environment around the junction itself. This approach, the so-called ``{\it environmental $P(E)$ theory}''~\cite{ENV}, will be considered separately below.

In order to estimate the rate $\Gamma_{\rm qp}$ of tunnel events from the superconductor to the nanowire, we start with a Fermi's golden rule approach, along the lines of Refs.~[\onlinecite{LutchynPhD,LarkinPRB2005,CatelaniGlazmanPRL2011,CatelaniGlazmanPRB2011}], and evaluate $\Gamma_{\rm qp}$ by averaging the matrix elements of the tunnel Hamiltonian over initial and final configurations with the appropriate thermal occupation factors:
\begin{equation}\label{eq:Fermi}
\Gamma_{\rm qp}=\frac{2\pi}{\hbar} \sum_{i,f} |\langle f|H_{\rm T}|i\rangle|^2 \delta\left( E_f -E_i \right)f(E_i)[1-f(E_f)]\;. 
\end{equation}
The initial and final states $|i\rangle$ and $|f\rangle$ are eigenstates of $H_0=H_{\rm SC}+H_{\rm TSC}$. The TSC state in the nanowire is induced by proximity effect, microscopically described by the same  Hamiltonian $H_{\rm T}$ that we are considering now. There is, however, no inconsistency, since the first order contribution Eq.~(\ref{eq:Fermi}) does not take into account the Cooper-pair hopping, which is assumed to be already implicitly included in $H_{\rm TSC}$.

We are interested in calculating matrix elements of the type
\begin{align}
\langle f|H_{\rm T}|i\rangle 
= \langle n_k=1,n_p=0| H_{\rm T} | n_k=0,n_p=1\rangle \;,
\end{align}
where we have indicated by $|n_k=0,n_p=1\rangle$ the product state with the TSC in its ground state (zero quasiparticles) and with one excess quasiparticle in the state $|p\rangle$ in the bulk superconductor. Correspondingly, $| n_k=1,n_p=0\rangle$ describes the state where the bulk SC is in its ground state, and one quasiparticle $|k\rangle$ is now present in the nanowire.
The matrix elements of the above equation can be evaluated by using the Bogoliubov transformation which diagonalizes the BCS Hamiltonian,
\begin{align}\label{Bogoliubov}
\nonumber a_{p,\sigma}^\dagger&=u_p \beta_{p,\sigma}^\dagger +\sigma v_p \beta_{-p,\bar\sigma} 
\\ a_{p,\sigma}&=u_p \beta_{p,\sigma} +\sigma v_p \beta_{-p,\bar\sigma}^\dagger \;.
\end{align} 
Here $u$ and $v$ are the usual particle-like and hole-like coherence factors, and $\bar\sigma=-\sigma$.
An analogous transformation can be applied to the $d_k$ operators in the nanowire, with corresponding $u/v$ coefficients~\footnote{For $p-$wave pairing the $u$ and $v$ amplitudes become 2$\times$2 matrices, but in our effective spinless case we assume we can still use scalar $s-$wave-like coefficients.}.
The explicit expression for the coherence factors is (we now denote them by $u_{\rm S,T}$ and $v_{\rm S,T}$ in order to make clear to which subsystem they refer to)
\begin{align}
u_{\rm S}^2(E),v_{\rm S}^2(E)=\frac{1}{2}\left[ 1 \pm \frac{\sqrt{E^2-\Delta^2_{\rm S}} }{E} \right]\;, 
\end{align}
and similarly for $u_{\rm T}(E)$ and $v_{\rm T}(E)$.
After this step the Hamiltonian $H_{\rm T}$ formally describes tunneling of quasiparticles $\beta_{p,\sigma}$ and $\eta_k$ across the junction. 
The insertion of the Bogoliubov transformation into the Fermi's golden rule produces the formula
\begin{align}\label{eq:Gamma2}
\nonumber \Gamma_{\rm qp}=&\frac{2\pi}{\hbar} \sum_{k,p,\sigma} |t_{kp\sigma}|^2  
\big[ u(E_p) u(E_k) -  v(E_p) v(E_k) \big]^2 \\
\nonumber  &\hspace{0,9cm} \times  f^{\rm neq}(E_p)[1-f(E_k)] ~\delta \left( E_k -E_p \right)\\
\simeq &\frac{2\pi \overline{|t|^2}}{\hbar} \sum_{k,p}  {\cal C}(E_k,E_p)  f^{\rm neq}(E_p)~\delta\left( E_k -E_p \right)\;.
\end{align}
Note that we have added a superscript to the Fermi occupancy factor in the bulk superconductor, to emphasize that its quasiparticles follow a non-equilibrium distribution, corresponding to the observed excess quasiparticle density. Nevertheless, $f^{\rm neq}(E)$ is still assumed to exhibit a sharp jump at $E=\Delta_{\rm S}$. Further considerations about  $f^{\rm neq}$ are developed in the following sections.
Assuming a weak energy- and momentum-dependence of the tunneling amplitude for energies close to the Fermi energy, we have extracted $t_{k,p}$ out of the summation and replaced it with an average squared tunneling amplitude $\overline{|t|^2}$. Further, we have made use of the low-temperature assumption to discard the term $f^{\rm neq}(E_p)f(E_k) $, since, as we will recall later in the paper, $\Delta_{\rm T}$ is typically only a fraction of  $\Delta_{\rm S}$, and $T\ll (\Delta_{\rm S}-\Delta_{\rm T})$.
Finally, the function ${\cal C}(E,E') \equiv \big[ u_{\rm S}(E) u_{\rm T}(E') - v_{\rm S}(E) v_{\rm T}(E') \big]^2$ has been introduced for brevity.

Converting the sum into integral and using the delta-function constrain gives us
\begin{align}\label{eq:GammaIntegral}
\nonumber \Gamma_{\rm qp}&\simeq \frac{\pi \overline{|t|^2}}{\hbar}\int_{\Delta_{\rm S}}^{\infty}{\rm d}E  \left[ 1- \frac{\Delta_{\rm S}\Delta_{\rm T}}{E^2 } \right]  2\nu_{\rm S}(E)2\nu_{\rm T}(E) f^{\rm neq}(E)
\\ 
&\simeq\frac{g_{\rm T}}{h} \!\!\bigintss_{\Delta_{\rm S}}^{\infty}\!\!\!\!\!\!  {{\rm d}E
\frac{
\left( E^2 - \Delta_{\rm S}\Delta_{\rm T} \right)}
{\sqrt{\left( E^2 -\Delta_{\rm S}^2\right) \left(E^2 -\Delta_{\rm T}^2 \right)}} }
f^{\rm neq}(E)\;.
\end{align}
The superconducting density of states $\nu(E)$ in the two subsystems is related to the normal-state density of states $\nu^{\rm n}(E)$ through the expression ($j={\rm S, T}$)
\begin{align}\label{eq:doss}
\frac{\nu_{j}(E)}{\nu^{\rm n}_j(E)}= \frac{E}{\sqrt{E^2-\Delta_j^2}}\;.
\end{align} 
The dimensionless tunneling conductance $g_{\rm T}$ is defined as ${h}/{(e^2R_{\rm T})}=R_{\rm Q}/R_{\rm T}$, with $R_{\rm Q}$ the quantum of resistance and $R_{\rm T}$
the normal-state resistance of the tunnel junction, determined by the formula
\begin{equation}\label{RT}
\frac{\hbar}{R_{\rm T}}={4\pi e^2}\sum_{k,p}|t_{k,p}|^2\delta(\xi_k)\delta(\xi_p)\simeq {4\pi e^2}\overline{|t|^2} \nu^{\rm n}_{\rm S}(0)\nu^{\rm n}_{\rm T}(0)\;.
\end{equation}
We would like to connect at this point the final expression for $\Gamma_{\rm qp}$ to the non-equilibrium density of quasiparticles, and use the experimentally measured values as an input for the calculation.
Using the assumption that $f^{\rm neq}(E)$ is exponentially peaked at $E=\Delta_{\rm S}$, we can approximate all the well-behaved factors in the integrand of Eq.~(\ref{eq:GammaIntegral}) by their value at $E=\Delta_{\rm S}$. 
Then, recalling the connection between the non-equilibrium quasiparticle density in the superconductor and the non-equilibrium Fermi distribution (from now on we will simply write $n_{\rm qp}^{\rm neq} =n^{\rm neq} $), 
\begin{equation}\label{f-n}
n^{\rm neq} =2\int {\rm d}E ~\nu_{\rm S}(E)f^{\rm neq}(E)\;,
\end{equation}
we can extract a factor $n^{\rm neq}$ from the integral Eq.~(\ref{eq:GammaIntegral}), and relate $\Gamma_{\rm qp}$ directly to the observed excess quasiparticle density~\cite{LutchynPhD,LarkinPRB2005,ShawPRB2008}:
\begin{equation}\label{eq:GammaNEQ-NOENV}
h\Gamma_{\rm qp}\simeq g_{\rm T}\frac{n^{\rm neq}}{2\nu^{\rm n}_{\rm S}} 
\sqrt{ \frac{\Delta_{\rm S}-\Delta_{\rm T}} {\Delta_{\rm S}+\Delta_{\rm T}} }
\;.
\end{equation}
The square root factor is of order unity for typical values of $\Delta_{\rm T}$.
Plugging in at this point the experimental values for $n^{\rm neq}\sim10/\mu{\rm m}^3$ (Ref.~[\onlinecite{ShawPRB2008}]), normal-state density of states $\nu^{\rm n}_{\rm S}\sim10^6/(\mu{\rm m}^3\cdot {\rm K})$,
and typical values for $g_{\rm T}$ in phase-qubit experiments ($R_{\rm T}\sim 10^2~\Omega \leftrightarrow g_{\rm T}\sim10^2$ in Refs.~[\onlinecite{MartinisPRL2009, nature08}]),
we obtain an estimation  for $\Gamma_{\rm qp}$ of the order of $\sim$~10~MHz. 

If instead one has higher tunnel resistances ($g_{\rm T}\sim1$ in Ref.~[\onlinecite{NaamanPRB06}] and $g_{\rm T}\sim10$ in Refs.~[\onlinecite{PeltonenPRB11,CorcolesAPL11}]), then the rate can be largely suppressed. Simply increasing the tunnel resistance however does not constitute a valid strategy in our situation, because also the tunneling of Cooper pairs would be reduced in that case, lowering the topological gap in the nanowire. For a more detailed discussion of this point, we refer the reader to the final section of the paper.

\section{Estimation with environmental $P(E)$ theory}
\begin{figure}[h!]
\centering
\includegraphics[width=1.0\linewidth]{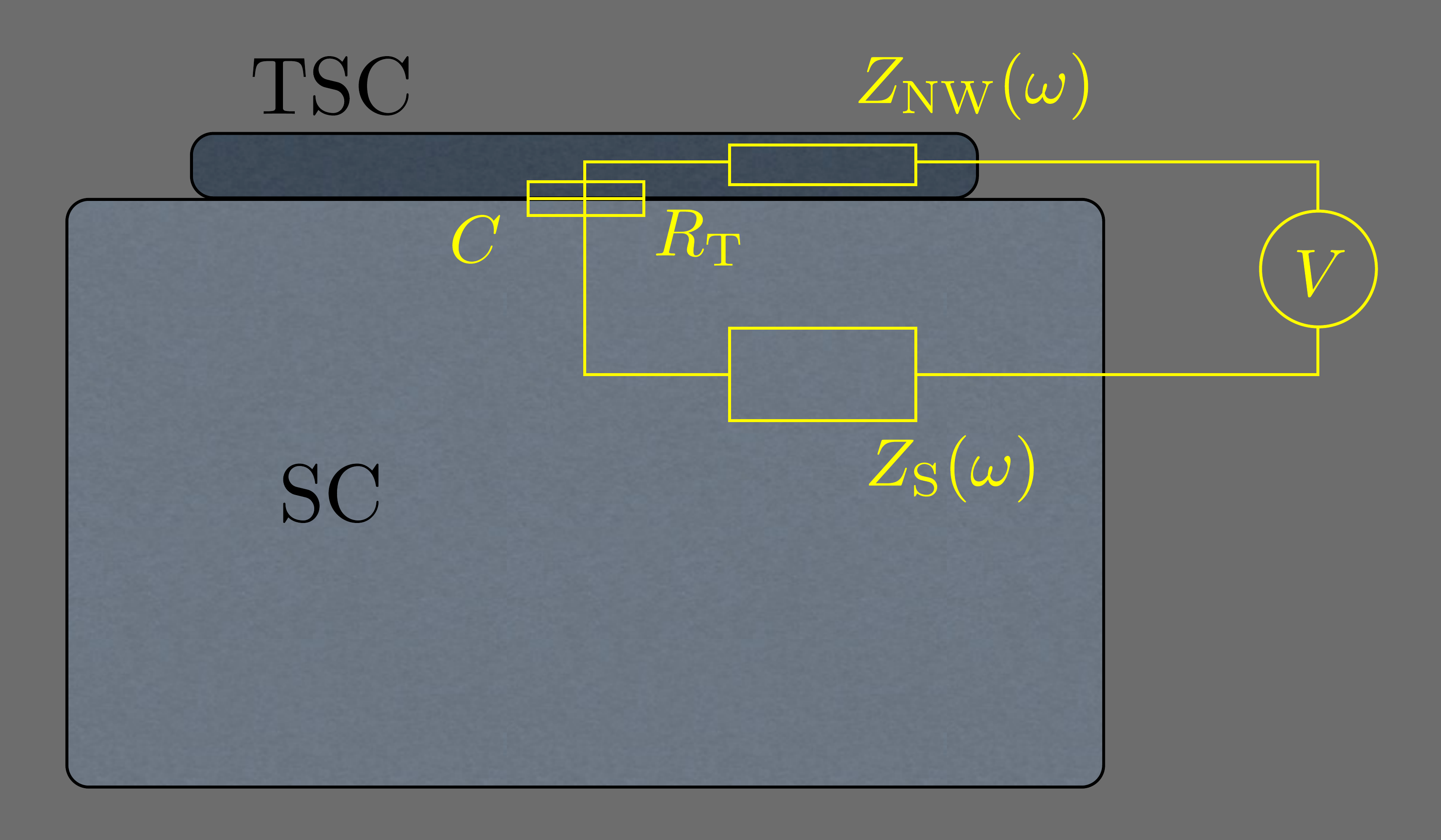}
\caption{\label{CircuitEQ} Schematic circuit representation of the Superconductor/Topological-nanowire system. The interface which separates the two subystems acts as a tunnel junction, with tunnel resistance $R_{\rm T}$ and capacitance $C$. The internal impedance $Z_{\rm S}$ and $Z_{\rm NW}$ of the superconductor and the nanowire are combined in the text in a single global environmental impedance $Z(\omega)$. An external voltage bias between the two sides of the junction can be present.}
\end{figure}

We now take into account the fact that the tunneling probability for a quasiparticle is influenced by the coupling with the surrounding environment, by making use of the environmental $P(E)$ theory~\cite{ENV}. This amounts to starting with the modified tunneling Hamiltonian
\begin{equation}
\tilde H_{\rm T}=\sum_{k,p,\sigma} \left( t_{kp\sigma} d^\dagger_{k} a_{p,\sigma} \right){\rm e}^{-i\varphi} +{\rm H.c.}\;,
\end{equation}
where the charge displacement operators ${\rm e}^{\pm i\varphi}$ act on the electrical circuit degrees of freedom (environment), and describe the transfer of a $\pm e$ charge through the SC/TSC junction in a tunneling event. Here, $\varphi$ is the conjugate coordinate to the charge $q$, with commutation relation $[q, \varphi] = ie$, and gives a charge displacement operator according to the relation ${\rm e}^{+ i\varphi}q{\rm e}^{- i\varphi} = q - e$.
Rewriting $H_{\rm T}$ in term of Bogoliubov operators, we obtain several terms, among which the ones describing the transfer of a quasiparticle have the form~\cite{MartinisPRL2009SI}
\begin{equation}
 \Big( u_{\rm S}u_{\rm T}{\rm e}^{-i\varphi} - v_{\rm S}v_{\rm T}{\rm e}^{i\varphi} \Big)
 \eta_k^\dagger \beta_{p,\sigma}\;.
\end{equation}
The evaluation of the modified tunnel rate 
\begin{equation}\label{HTtilde}
\sum_{i,f} |\langle f| \tilde H_{\rm T}|i\rangle|^2 \delta\left( E_f -E_i \right)f^{\rm neq}(E_i)[1-f(E_f)]
\end{equation}
now involves also averages over environment degrees of freedom, and it requires the calculation of the equilibrium correlation function
\begin{equation}
\left\langle  {\rm e}^{i\varphi(t)} {\rm e}^{-i\varphi(0)}\right\rangle = 
{\rm e}^{\left\langle  \left[ \varphi(t)-\varphi(0)\right]\varphi(0)\right\rangle}
\equiv {\rm e}^{J(t)}\;,
\end{equation}
which in the case of Bogoliubov-quasiparticle tunneling must be properly corrected, as explained by Martinis {\it et al.}~\cite{MartinisPRL2009SI}, and becomes
\begin{align}\label{eJ(t)tilda}
{\rm e}^{\tilde J(t)} &= 
 \bigg[(u^2+v^2) {\rm e}^{\left\langle  \varphi(t)\varphi(0)\right\rangle}
-2uv {\rm e}^{-\left\langle  \varphi(t)\varphi(0)\right\rangle} \bigg] \nonumber \\
&\times  {\rm e}^{-\left\langle  \varphi(0)\varphi(0)\right\rangle}\;.
\end{align}
Here for sake of brevity we wrote $u=u_{\rm S}u_{\rm T}$ and $v=v_{\rm S}v_{\rm T}$.
The fluctuation-dissipation theorem provides us with a relation between the correlation function $J(t)$ and the dissipation in the environment, indirectly described by its impedance~\cite{ENV}:
\begin{align}\label{J(t)}
J(t)&= \left\langle  \left[ \varphi(t)-\varphi(0)\right]\varphi(0)\right\rangle \nonumber \\
&=2\int_{-\infty}^{\infty} \frac{d\omega}{\omega}
\frac{\re Z_{\rm t}(\omega)}{R_{\rm Q}}  (e^{-\textrm{i}\omega t}-1)\;,
\end{align}
where $Z_{\rm t}(\omega)$ is the total environmental impedance, $R_{\rm Q}=h/e^2$ is the quantum of resistance for single-electron charge transfer, and we have assumed $T=0$ (while it is still necessary to use a finite value of $T$ in the tunnel rate calculations). 
This description in terms of circuitry elements, where the tunnel junction is characterize by its capacitance $C$ and tunnel resistance $R_{\rm T}$, and the environment properties are encoded in its impedance, is summarized in Fig.~\ref{CircuitEQ}.
Due to the presence of the delta function in the summations of Eqs.~(\ref{eq:Fermi}) and (\ref{HTtilde}), what we finally need is the Fourier transform of $J(t)$ (and $\tilde J(t)$ respectively), usually named $P(E)$:
\begin{align}\label{P(E)}
\int_{-\infty}^{\infty} \frac{dt}{2\pi\hbar} {\rm e}^{J(t)} {\rm e}^{\textrm{i}E t/\hbar } \equiv P(E) \;.
\end{align}
In terms of such function, the tunneling rate in the case of electron-environment coupling is expressed as
\begin{align}\label{eq:GammaENV}
\Gamma_{\rm qp}=&\frac{4\pi \overline{|t|^2}}{\hbar}  \int_{\Delta_{\rm S}}^{\infty} {\rm d}E \int_{\Delta_{\rm T}}^{\infty}  {\rm d}E'  \times  \\
& \nu_{\rm S}(E)\nu_{\rm T}(E') f^{\rm neq}(E)[1-f(E')] P(E-E') \nonumber\;,
\end{align}
where now $P(E-E')$ may be interpreted as the probability of a tunnel event which involves an energy exchange $(E-E')$ between quasiparticle and environmental degrees of freedom (to be precise, the energy $E-E'$ is the energy transferred from the tunneling particle to the environment).

Going back now to the modified correlation function $\tilde J(t)$,  few comments are in order. The first term of the rhs of Eq.~(\ref{eJ(t)tilda}) equals $(u^2+v^2){\rm e}^{J(t)}$, and the same steps described above lead to contribution $(u^2+v^2)P(E)\equiv{\cal C}_+(E)P(E)$. 
The correlator $\left\langle \varphi(t)\varphi(0)\right\rangle$ is not well defined, due to an infrared divergence, for impedances whose real part does not vanish at $\omega=0$, see Eq.~(\ref{J(t)}). This does not constitute a problem for the physical quantity $J(t)$, since there the diverging static correlation $\varphi^2$ is subtracted off.
The same does not happen with the second term, which instead involves the factor ${\rm e}^{-\left\langle  \varphi(t)\varphi(0)\right\rangle}$, not compensated by ${\rm e}^{-\left\langle  \varphi(0)\varphi(0)\right\rangle}$. Since $\left\langle  \varphi(t)\varphi(0)\right\rangle$ is positively diverging, however, this second term in Eq.~(\ref{eJ(t)tilda}) vanishes. 
In Ref.~[\onlinecite{MartinisPRL2009SI}] this issue is not present, since they consider a model where $\re Z_{\rm t}(0)=0$, and the divergence is absent. 
%

\subsection{Single-mode environment}
\begin{figure}[h!]
\centering
\includegraphics[width=0.7\linewidth]{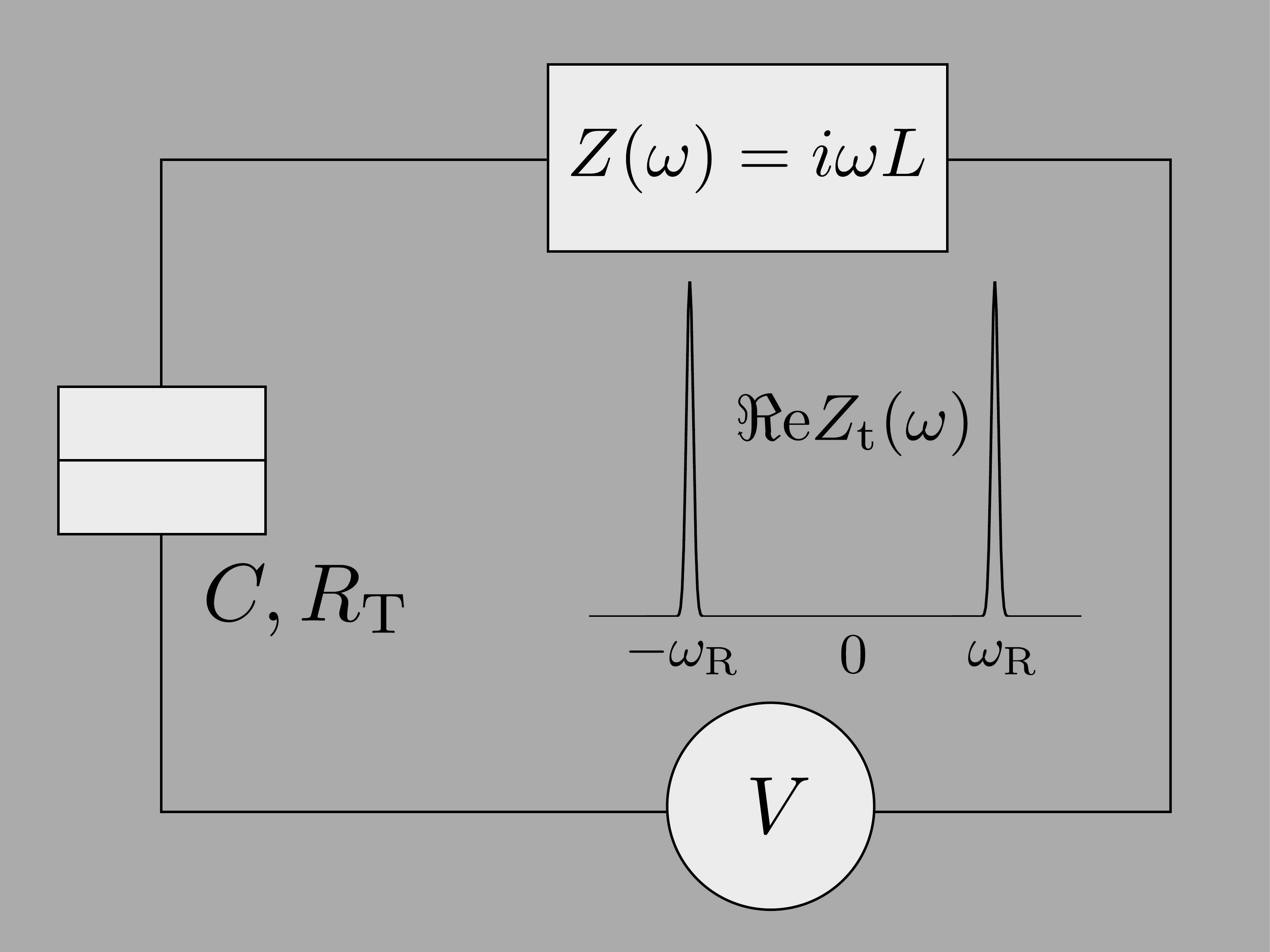}
\caption{\label{Circuit-SM} Equivalent circuit for the SC/TSC system in the case of a single-mode environment. The environment is modeled with a single inductance $L$ with impedance $Z(\omega)=i\omega L$, corresponding to a total impedance $\re \left[ Z_{\rm t}(\omega)\right] \sim  \left[\delta(\omega+\omega_{\rm R})+\delta(\omega-\omega_{\rm R})\right]$ (plotted in the inset) .}
\end{figure}

One can get the simplest model for the environment by studying the coupling of the tunnel junction to one single environmental mode, which could come from a resonance in the lead impedance $Z(\omega)$ of from bound states in the barrier.
Such coupling can be implemented  by putting an inductor with inductance $L$ into the external circuit. Seen from the junction, the impedance $Z(\omega)=i\omega L$ is in parallel with the capacitance $C$ of the junction itself, and the total impedance reads~\cite{ENV}
\begin{align}
Z_{\rm t}(\omega) = \frac{1}{i\omega C +Z^{-1}(\omega)}=\frac{1}{C}\frac{i\omega}{\left[ \omega_{\rm R}^2-(\omega-i\epsilon)^2\right]}\;,
\end{align}
with environmental resonance frequency $\omega_{\rm R}=1/\sqrt{LC}$. The infinitesimal imaginary part $\epsilon$ is necessary in order to obtain the correct result for the real part of the impedance. By taking the limit $\epsilon\rightarrow0$ one gets~\cite{ENV, MartinisPRL2009}
\begin{align}
\re \left[ Z_{\rm t}(\omega)\right] = \frac{\pi}{2C} \left[\delta(\omega+\omega_{\rm R})+\delta(\omega-\omega_{\rm R})\right]
\label{eqZ} \;.
\end{align}
This expression is essentially saying that the resonator can both absorb or emit photons (mode quanta) at frequency $\omega_{\rm R}$. In our case, if we identify the environmental mode with the only available low-energy excitation in the nanowire-superconductor system, \ie the Majorana mode, we obtain a resonance energy $\hbar\omega_{\rm R}\simeq0$ (or energy much smaller than all other energy scales). This situation is sketched in Fig.~\ref{Circuit-SM}.

Before inserting this form of the total impedance in the formula Eq.~(\ref{P(E)}) for $P(E)$, let us define the parameter~\cite{ENV}
\begin{equation}\label{zeta}
\zeta\equiv \frac{\pi}{\omega_{\rm R}}\frac{1}{R_{\rm Q}C}=\frac{E_{\rm C}}{\hbar \omega_{\rm R}}
\end{equation}
which compares the single-electron charging energy with the environmental mode excitation energy.
This parameter determines the size of charge fluctuations~\cite{ENV}:
\begin{equation}
\langle Q^2 \rangle=\frac{e^2}{4\zeta}\coth\left( \frac{\hbar \omega_{\rm R}}{2T} \right) \;
\end{equation}
Using the definition of $\zeta$, the expression for $P(E)$ in the single-mode limit becomes
\begin{align}
P(E) &= \int_{-\infty}^{\infty} \frac{dt}{2\pi\hbar} e^{\textrm{i}E t/\hbar }
\exp\left[  \zeta \left(e^{-\textrm{i}\omega_r t}-1\right)\right]  \label{eqP1} \; .
\end{align}
In typical superconducting charge-qubit experiments one has~\cite{MartinisPRL2009} that $E_{\rm C}\ll \hbar \omega_{\rm R}$, that is $\zeta\ll1$, and then the external exponential in Eq.~(\ref{eqP1}) can be expanded around zero. In our case instead, since $\omega_{\rm R}\sim0$, such simplification is not possible, and we rather expand the internal exponential for $t\ll \hbar/\omega_{\rm R}$. The result is:
\begin{align}\label{P(E)-single}
J(t) &\simeq \zeta\left(-i \omega_{\rm R} t-\omega^2_{\rm R}t^2\right) =-\frac{E_{\rm C}}{\hbar} \left( it + \omega_{\rm R}t^2 \right)\\
P(E)&\simeq \frac{{\exp}\left[{-\frac{(E-E_{\rm C})^2}{4E_{\rm C}\hbar\omega_{\rm R}}}\right] }{\sqrt{4\pi E_{\rm C}\hbar \omega_{\rm R}}}\xrightarrow[~~\omega_{\rm R}\rightarrow0~~]{}
\delta\left( E-E_{\rm C}\right)\; .
\end{align}
That is, in first approximation the energy exchange between quasiparticles and environment occurs with unit probability and is peaked at the charging energy $E_{\rm C}=e^2/(2C)$.
In the opposite limit $\zeta\ll1$ valid for typical superconducting qubits~\cite{ShawPRB2008,MartinisPRL2009}, one would get instead that the energy exchange is peaked at the resonator energy $\omega_{\rm R}$, and that the probability $\sim\zeta$ for such exchange is very small (the most probable event being the tunneling of a quasiparticle without energy flow to the environment).

Plugging now into Eq.~(\ref{eq:GammaENV}) the form of $P(E)$ just obtained, we get
\begin{align}\label{eq:GammaENV2}
&h\Gamma_{\rm qp}=\nonumber \\
&{g_{\rm T}} \bigintss_{\Delta_{\rm S}}^{\infty} \!\!\!\!{\rm d}E ~
\frac{
\left[ E\left(E+E_{\rm C}\right) - \Delta_{\rm S}\Delta_{\rm T} \right]}
{\sqrt{\big[ \left(E+E_{\rm C}\right)^2 -\Delta_{\rm T}^2\big] \left[E^2 -\Delta_{\rm S}^2 \right]}}
f^{\rm neq}(E)\;.
\end{align}
This is essentially identical to the previous result Eq.~(\ref{eq:GammaIntegral}), with the simple substitution $E_p \rightarrow (E_p+E_{\rm C})$, and leads to the low-temperature result
\begin{align}\label{eq:GammaSM}
h\Gamma_{\rm qp}\approx  g_{\rm T}\frac{n^{\rm neq}}{2\nu^{\rm n}_{\rm S}} 
 \sqrt{\frac{\Delta_{\rm S}-\Delta_{\rm T}+E_{\rm C}}{\Delta_{\rm S}+ \Delta_{\rm T}+E_{\rm C}}}
\;.
\end{align}
Compared to equation (\ref{eq:GammaNEQ-NOENV}), the presence of $E_{\rm C}$ produces a negligible modification to the quantitative estimate for $\Gamma_{\rm qp}$ in the case of typical values for $C$ ($\sim1$~pF) and $E_{\rm C}\sim 0.1~\mu$eV $\ll \Delta_{\rm S},\Delta_{\rm T}$. Thus, even in this case the typical tunneling rate turns out to be $\Gamma_{\rm qp}\sim$ 100 kHz$-$10 MHz, depending on the transparency of the tunnel barrier.\\

\subsection{Ohmic environment}
\label{sec:ohmic}
\begin{figure}[h!]
\centering
\includegraphics[width=0.7\linewidth]{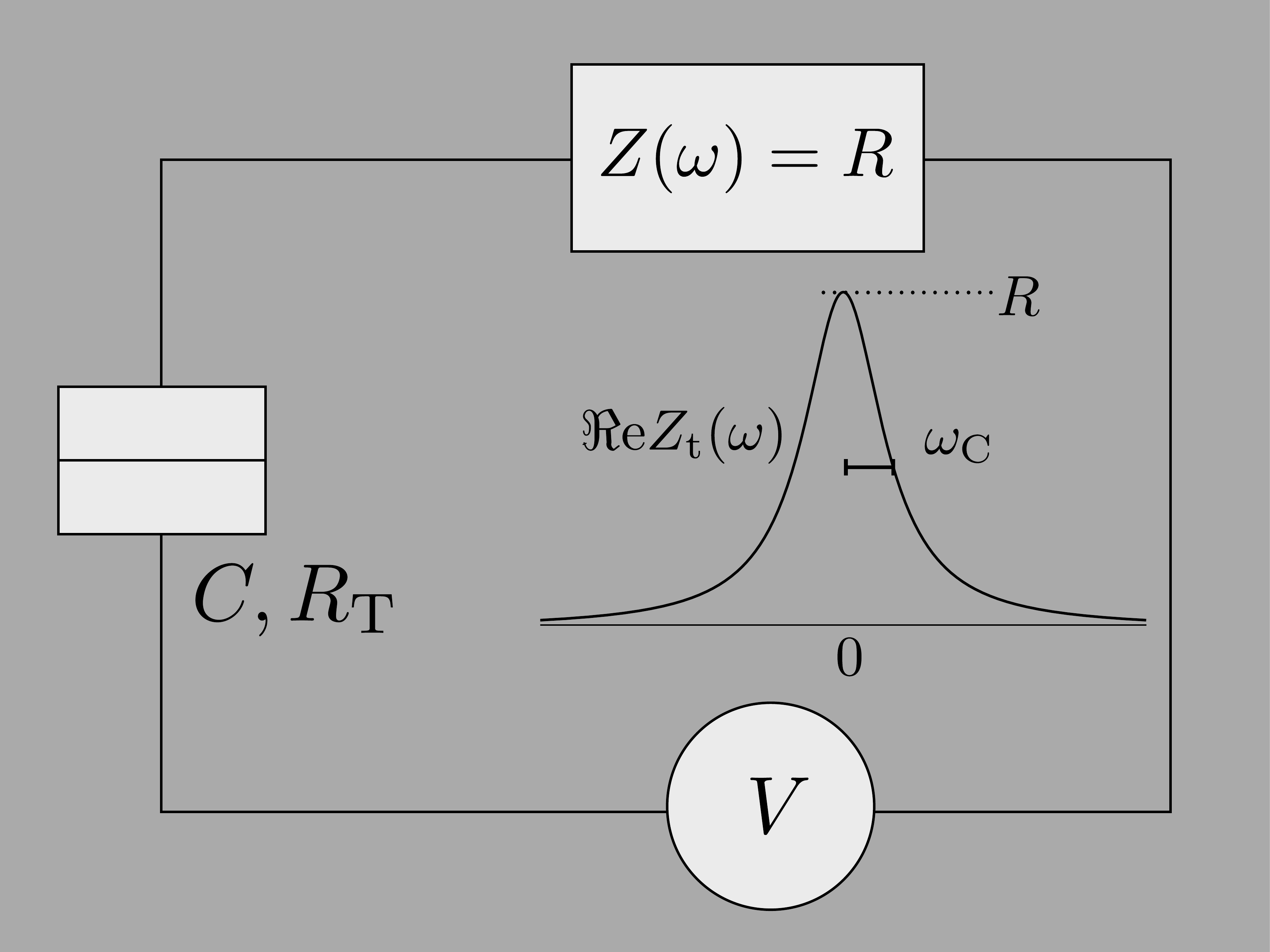}
\caption{\label{Circuit} Same as in Fig.~(\ref{Circuit-SM}), but with a different environmental impedance. Here the environment is modeled by $Z=R$, and the resulting Lorentzian total impedance is shown in the inset, where $\omega_{\rm C}= {1}/{(RC)} ={gE_{\rm C}}/{\pi\hbar}$.}
\end{figure}

We consider now the more realistic case of an Ohmic environment, with external impedance $Z(\omega)=R$ and total impedance
\begin{align}
\nonumber \frac{\re \left[ Z_{\rm t}(\omega)\right]}{R_{\rm Q}}=\frac{1}{R_{\rm Q}}\re \left[\frac{1}{i\omega C +1/R}\right]=
\frac{1}{g}\frac{1}{\big[1+\left(\omega/\omega_{\rm C}\right)^2\big]}\,,
\end{align}
where we have introduced the dimensionless environmental conductance $g\equiv R_{\rm Q}/R$ and the 
frequency 
\begin{align}
\omega_{\rm C}\equiv \frac{1}{RC} =\frac{g}{\pi}\frac{E_{\rm C}}{\hbar}\;,
\end{align}
which represents an effective cutoff for the total impedance, due to the junction capacitance: at energies small compared to $\hbar\omega_{\rm C}$ the real part of the total impedance is essentially given by $R$, while for higher energies $Z_{\rm t}(\omega)$ decreases. Such behavior is shown in Fig.~\ref{Circuit}.
The $P(E)$ corresponding to this case cannot be calculated analytically, but the low-energy and the high-energy asymptotic behaviors can be obtained as~\cite{ENV}:
\begin{eqnarray}
P(E)= \left\{ \begin{array}{crc}
{\displaystyle  \frac{{\rm e}^{-2\gamma/g}}{\Gamma(2/g)}\frac{1}{E}\left( \frac{\pi}{g} \frac{E}{E_{\rm C}} \right)^{2/g} \quad}  &{\rm for~}& E\ll E_{\rm C}\;, \\
\\
{\displaystyle \frac{2g}{\pi^2}\frac{E_{\rm C}^2}{E^3}} \quad &{\rm for~}& E\gg E_{\rm C} \;.
\end{array}\right.
\end{eqnarray}
Here $\gamma$ is the Euler constant and $\Gamma$ the gamma-function. The behavior of $P(E)$ for intermediate energies has to be evaluated numerically. Since we are mostly interested in energy exchanges between the superconductor and the topological nanowire of the order of $\delta\Delta\equiv(\Delta_{\rm S}-\Delta_{\rm T})\sim{\cal O}(\Delta_{\rm S})\sim $ meV, and since most typically $E_{\rm C}\ll\Delta_{\rm S}$, we are not justified to use the small-energy expansion of $P(E)$, and we must rather determine $P(E)$ numerically. 
\begin{figure}[h!]
\centering
\includegraphics[width=1.0\linewidth]{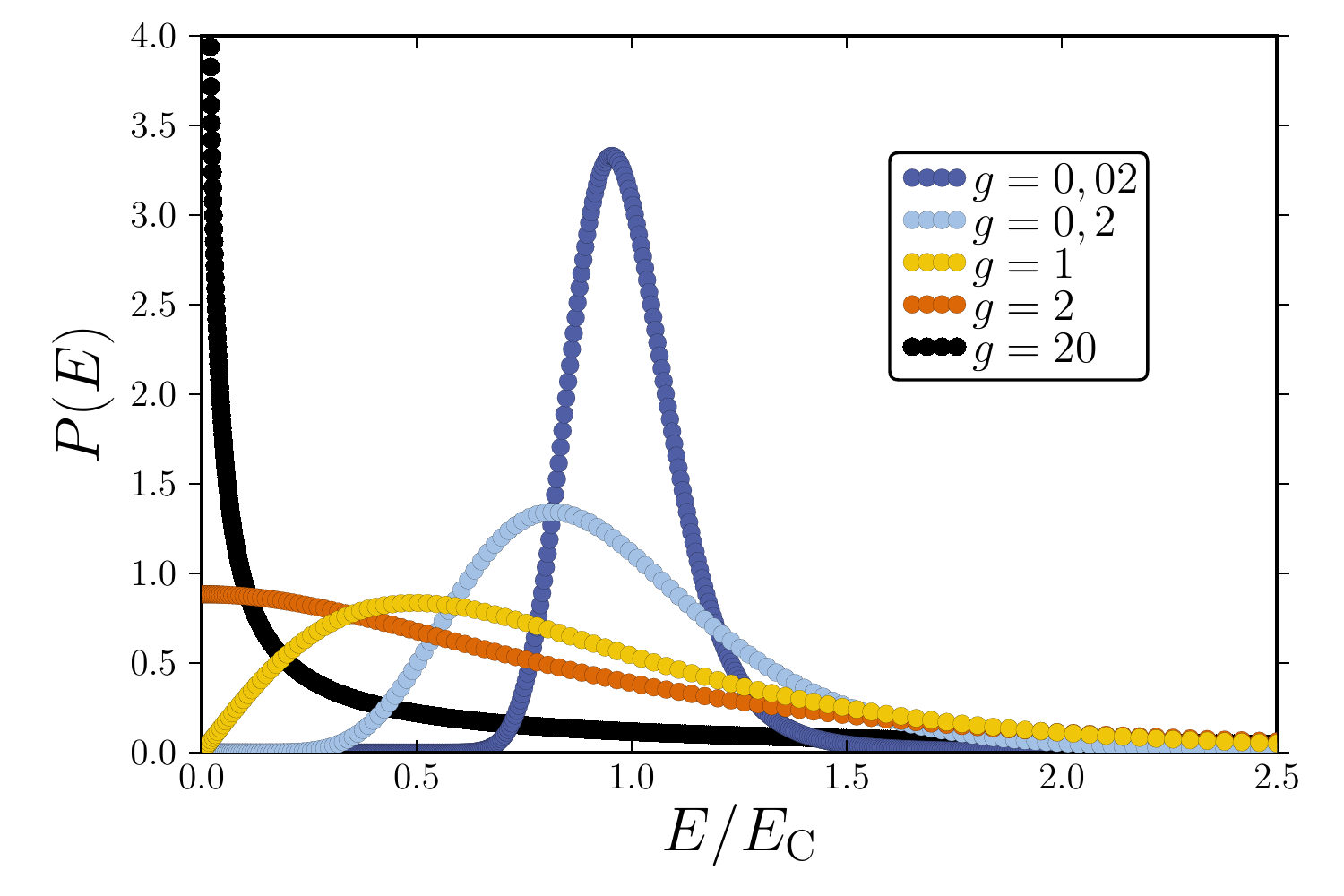}
\caption{\label{fig:P(E)} Behavior of the probability function $P(E)$ as a function of the energy exchange measured in units of $E_{\rm C}$. For large environment resistances (small $g$) the junction releases a typical energy amount of the order of the charging energy. For small resistances (large $g$) the energy which is exchanged shrinks to zero, and one recovers a situation with independent quasiparticles and junction degrees of freedom. The curves have been obtained through numerical integration.}
\end{figure}
By taking the derivative  of Eq.~(\ref{J(t)}) and performing a Fourier transform, one finds that $P(E)$ obeys to the integral equation
\begin{align}\label{IEforP}
EP(E)=\frac{2}{R_{\rm Q}}\int_0^E \!{\rm d}E' ~\re\left[ Z_{\rm t}\left( \frac{E-E'}{\hbar}\right)\right] P(E')\;,
\end{align}
which can be solved for example by iteration. A collection of solutions for different values of the parameter $g$ is shown in Fig.~\ref{fig:P(E)}. Qualitatively different behaviors are observed in the highly resistive and low-resistive limits.
Inserting the obtained solution $P(E)$ into Equation (\ref{eq:GammaENV}) we can get the desired estimation for $\Gamma_{\rm qp}$ in this case. However, due to the finite energy exchange allowed by $P(E)$, the singularities of the two density of states distributions can overlap in the integral, and caution must be exercised. In particular, one cannot always make the simplification adopted to attain Eq. (\ref{eq:GammaNEQ-NOENV}), which allowed us to single out a factor $n^{\rm neq}$. 

\begin{figure}[h!]
\centering
\includegraphics[width=1.0\linewidth]{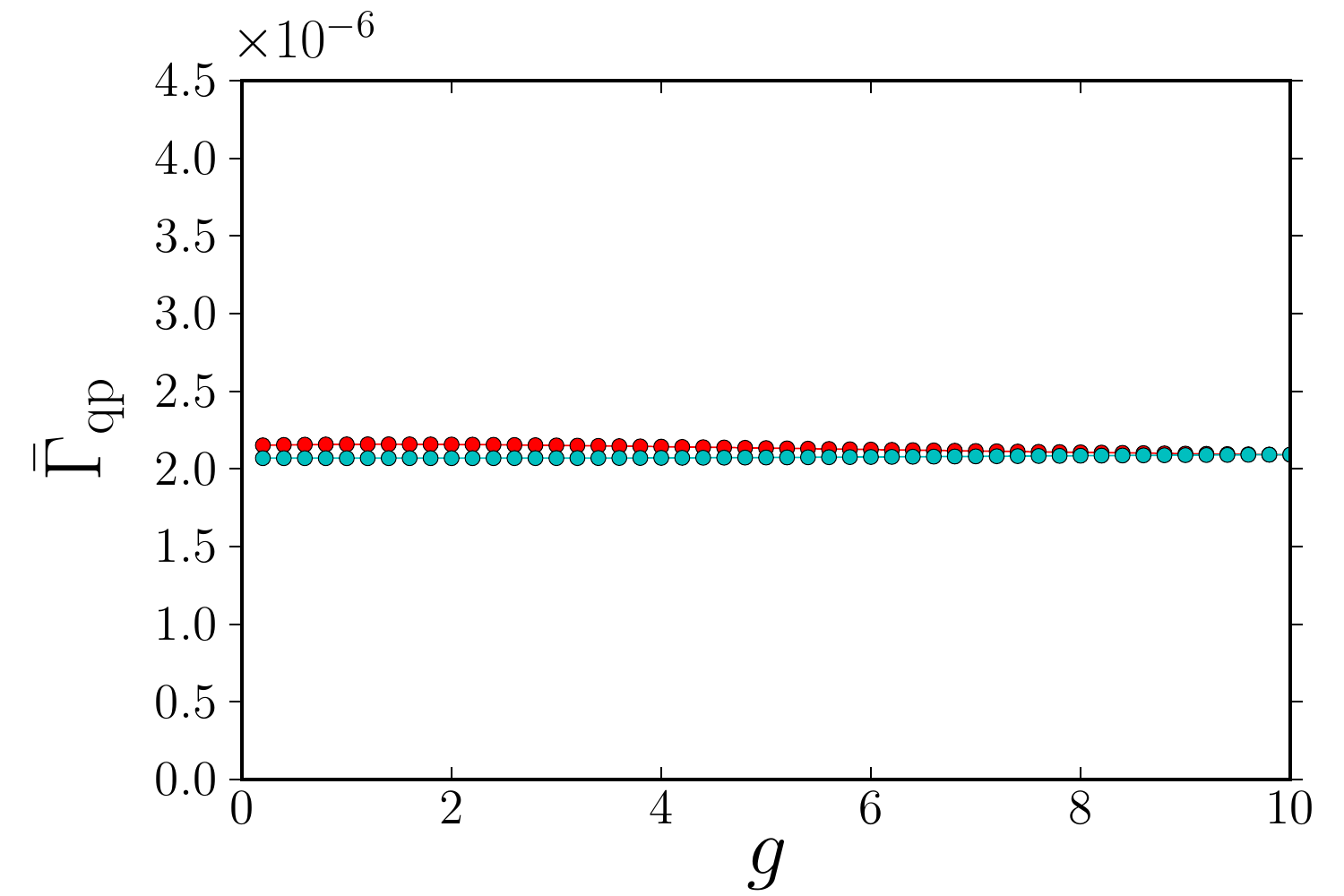}
\caption{\label{fig:GammaA} Dimensionless quasiparticle tunneling rate as a function of environmental dimensionless conductance $g$. The red curve refers to the case $\Delta_{\rm S}=10E_{\rm C}$, $\Delta_{\rm T}=5E_{\rm C}$, and no observable dependence on $g$ is noticed at this scale. The cyan curve has been obtained for $\Delta_{\rm S}=100E_{\rm C}$, $\Delta_{\rm T}=50E_{\rm C}$, and the corresponding values of $\Gamma_{\rm qp}$ are slightly lower in this case. }
\end{figure}

In the case $E_{\rm C}\ll \delta\Delta$ the same approximation can still be safely employed, since the probability distribution $P(E)$ is appreciably different from zero only in a support $\sim[0: E_{\rm C}]$, for all values of $g$ [see Fig.~(\ref{fig:P(E)})], and the two singularities in the densities of states $\nu_{\rm S}$ and $\nu_{\rm T}$, located at $\Delta_{\rm S}$ and $\Delta_{\rm T}$ respectively, overlap only through the high-energy tail of $P(E)$, without significant contributions to the integral for $\Gamma_{\rm qp}$. We then get
\begin{align}
\nonumber
\int_{\Delta_{\rm S}} \int_{\Delta_{\rm T}} \!\!\!{\rm d}E {\rm d}E' ~\nu_{\rm S}(E)f^{\rm neq}(E)
 \nu_{\rm T}(E')\tilde P(E,E')\approx \\ 
\int_{\Delta_{\rm S}} \!\!\!{\rm d}E ~\nu_{\rm S}(E)f^{\rm neq}(E) \cdot \int_{\Delta_{\rm T}} \!\!\!{\rm d}E' \nu_{\rm T}(E')\tilde P(\Delta_{\rm S},E') \nonumber \\
\propto  n^{\rm neq} \int_{\Delta_{\rm T}} \!\!\!{\rm d}E' \nu_{\rm T}(E')\tilde P(\Delta_{\rm S},E') \;,
\end{align}
where we used the notation $\tilde P(E,E')={\cal C}(E,E')P(E-E')$.
The resulting $\Gamma_{\rm qp}(g,\Delta_{\rm S},\Delta_{\rm T})$ is shown as a function of $g$ for some specific choices of $\Delta_{\rm S}$ and $\Delta_{\rm T}$ in Fig.~\ref{fig:GammaA}. We choose to plot the dimensionless quantity $\bar \Gamma_{\rm qp}\equiv h\Gamma_{\rm qp}/(g_{\rm T}\Delta_{\rm S})$, meaning that the quasiparticle tunnel rate is measured in units of $\Delta_{\rm S}/h$, and has been divided by $g_{\rm T}$. A superconducting gap of 2 K corresponds to a frequency of 40 GHz, and for $g_{\rm T}=10^2$ the values shown in the figure indicate then $\Gamma_{\rm qp}\simeq$ 10 MHz.

In the more interesting case $E_{\rm C}\gtrsim \delta\Delta$ (realized for example for $\Delta_{\rm S}\simeq 100~\mu$eV and $C\simeq1$fF) the environment can couple energy regions where the singularities in the density of states of the two subsystems occur. Now the approximations adopted above are not justified anymore, especially for small values of $g$, and one needs in principle to solve the full two-dimensional integral in Eq.~\ref{eq:GammaENV}. The problem then is, without the decoupling of the integrals we cannot extract anymore a factor $n^{\rm neq}$. We then need an explicit estimate for the unknown term $f_{\rm S}^{\rm neq}$. This can be done by assuming that the quasiparticles, while still being in thermal equilibrium at temperature $T$, are out of electro-chemical equilibrium, and the excess quasiparticle density $n^{\rm neq}$ can be accounted for by an effective chemical potential shift $\tilde\mu$:
\begin{align}
n^{\rm neq}=\int_{\Delta_{\rm S}} \!\!{\rm d}E ~\nu_{\rm S}(E)\frac{1}{\left[{\rm e}^{(E-\tilde\mu)/T}+1\right]}\;.
\end{align}
To lowest order in temperature, we can connect $\tilde\mu$ directly to $n^{\rm neq}$ as~\cite{ShawPRB2008}
\begin{equation}
\tilde\mu\simeq T\ln \left(\frac{n^{\rm neq}}{n^{\rm eq}} \right) \;.
\end{equation}

\begin{figure}[h!]
\centering
\includegraphics[width=1.0\linewidth]{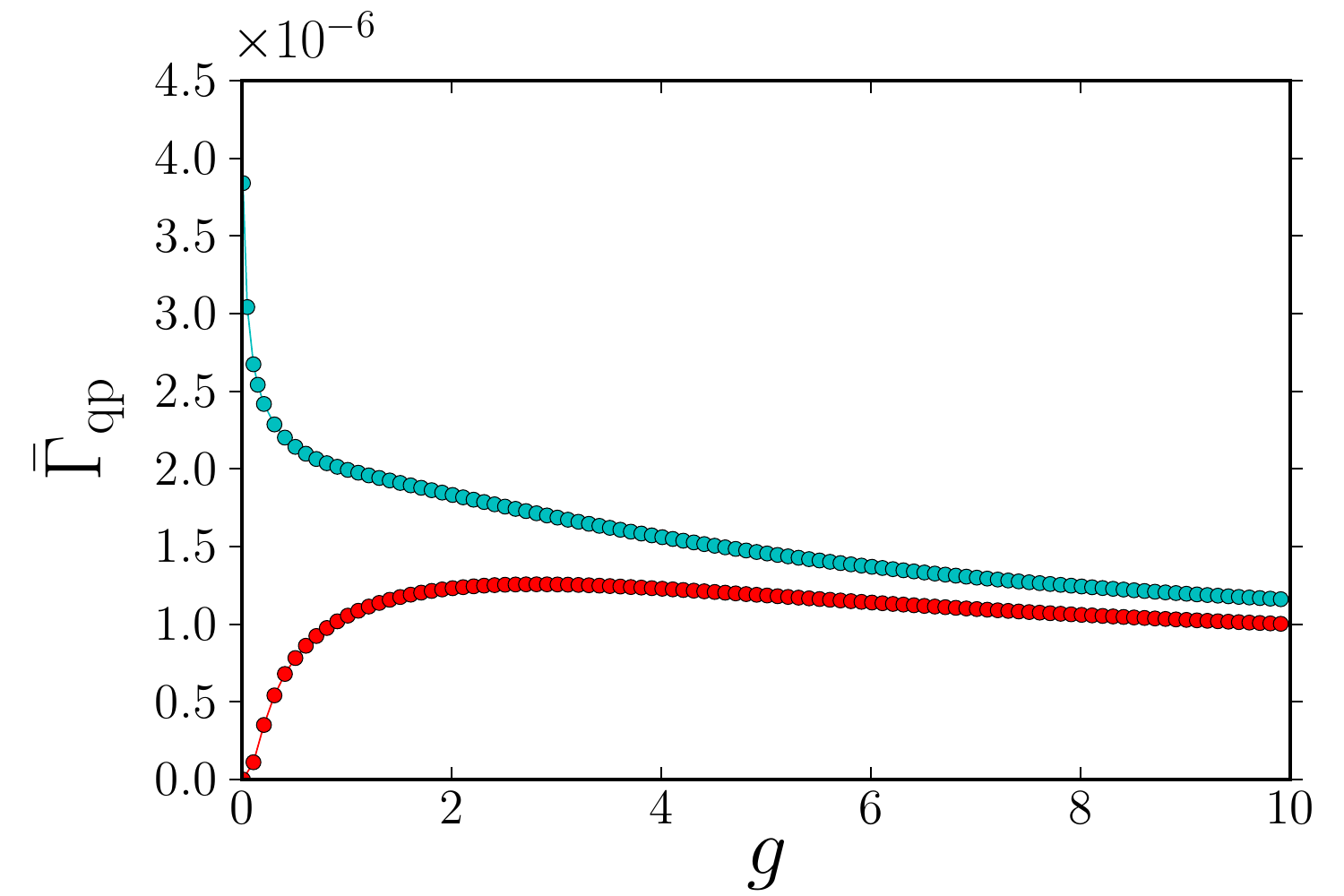}
\caption{\label{fig:GammaB} Same as in Fig.~\ref{fig:GammaA}, but with different parameter values. The cyan curve corresponds to $\Delta_{\rm S}=2E_{\rm C}$, $\Delta_{\rm T}=E_{\rm C}$ and the red curve refers to the case $\Delta_{\rm S}=E_{\rm C}$, $\Delta_{\rm T}=0.5E_{\rm C}$. 
In the first case, $(\Delta_{\rm S}-\Delta_{\rm T})$ equals $E_{\rm C}$ and leads to an unbounded increase in $\Gamma_{\rm qp}$ for $g\rightarrow0$. 
In the second case (and in general for $E_{\rm C}>\delta\Delta$) one observes $\Gamma_{\rm qp}(g\rightarrow0)\rightarrow0$ because the energy exchange $E_{\rm C}$ provided by the environment is too large to be absorbed by $\delta\Delta$.}
\end{figure}
Inserting the calculated $\tilde\mu$ in the formula for $\Gamma_{\rm qp}$ and performing the double integration, one can get numerical estimations for any value of the parameters $\Delta_{\rm S}/E_{\rm C}$ and $\Delta_{\rm T}/E_{\rm C}$. In Fig.~\ref{fig:GammaB} we report (cyan curve) the results for the ``worst'' case $(\Delta_{\rm S}-\Delta_{\rm T})=E_{\rm C}$. 
{One can see that in the limit $g\rightarrow0$ the quasiparticle poisoning rate is strongly enhanced, due to the perfect coupling of the two singularities in the density of states}. However, this regime is difficult to attain, and the strong increase in $\Gamma_{\rm qp}$ is localized at $g\simeq0$ which requires unrealistic environmental resistances $R\gg R_{\rm Q}.$ In conclusion then, this issue should not represent a problem.

In the regime $(\Delta_{\rm S}-\Delta_{\rm T})<E_{\rm C}$ (red curve), the environment provides for $g\lesssim2$  (see Fig.~\ref{fig:P(E)}) a typical energy larger than the ``energy distance" between the two subsystems, and since $\nu_{\rm T}(E<\Delta_{\rm T})=0$, smaller values for $\Gamma_{\rm qp}$ are obtained for decreasing $g$. In the limit $g \rightarrow0 $ we have $P(E)\propto\delta(E-E_{\rm C})$ and the result of integration is suppressed to zero. 
Note that for $g \rightarrow\infty$ the two curves of Fig.~\ref{fig:GammaB} approach each other, because in that limit $P(E)$ is peaked in $E=0$ and the exact position of $\Delta_{\rm T}$ with respect to $\Delta_{\rm S}$ becomes irrelevant.

\section{Quantitative considerations}
The final estimations strongly depend on the value of the tunneling resistance $R_{\rm T}$ which enters the expression for the poisoning rate. As anticipated above, such values are different for different experiments, ranging from $\sim10~\Omega$ to $\sim 10^4~\Omega$. 
By looking at the expression for the quasiparticle tunnel rate, Eqs.~(\ref{eq:GammaNEQ-NOENV}) and (\ref{eq:GammaSM}) 
, one could conclude that large tunnel resistances (low $g_{\rm T}$) are desirable so that $\Gamma_{\rm qp}$ is reduced. But as we already commented, by the same token also Cooper-pair tunneling would be suppressed, and hence the proximity-induced gap would get reduced. Analytical calculations~\cite{proximity} have shown that the pairing potential amplitude $\Delta_{\rm pr}$ induced in the proximized system, in terms of the parental pairing amplitude $\Delta_{\rm S}$ and of the microscopic tunneling rate $\Gamma_0$, is given by
\begin{align}\label{DeltaPR}
\Delta_{\rm pr}=\frac{\Gamma_0 }{\Gamma_0+\Delta_{\rm S}}\Delta_{\rm S}\;.
\end{align}
The tunneling rate for bare electrons is evaluated as
\begin{align}\label{Gamma0}
\Gamma_0=\pi\overline{|t|^2} \nu^{\rm n}_{\rm S}(0)\;,
\end{align}
so that, using the definition Eq.~(\ref{RT}) for $R_{\rm T}$, one can relate $\Gamma_0$ and $R_{\rm T}$ as:
\begin{align}\label{Gamma0-RT}
\Gamma_0=\frac{R_{\rm Q}}{8\pi R_{\rm T}}\frac{1}{\nu^{\rm n}_{\rm T}(0)}\;.
\end{align}
In the low transparency limit, $\Gamma_0\ll\Delta_{\rm S}$, the proximity gap is set by $\Gamma_0$, see Eq.~(\ref{DeltaPR}), and is therefore rather small.
On top of that, the topological gap is further reduced due to the Rashba and Zeeman interaction:
\begin{align}\label{DeltaT}
\Delta_{\rm T}=\frac{\alpha k_{\rm F}}{\sqrt{V_{\rm Z}^2+(\alpha k_{\rm F})^2}}\Delta_{\rm pr}=\frac{1}{\sqrt{1+\chi^2}}\Delta_{\rm pr}\;,
\end{align}
with $\chi\equiv V_{\rm Z}/(\alpha k_{\rm F})$ quantifying the ratio between Zeeman splitting and typical spin-orbit interaction. Note that one always has $\Delta_{\rm T}\leq\Delta_{\rm pr}\leq\Delta_{\rm S}$.
\begin{figure}[h!]
\centering
\includegraphics[width=1.0\linewidth]{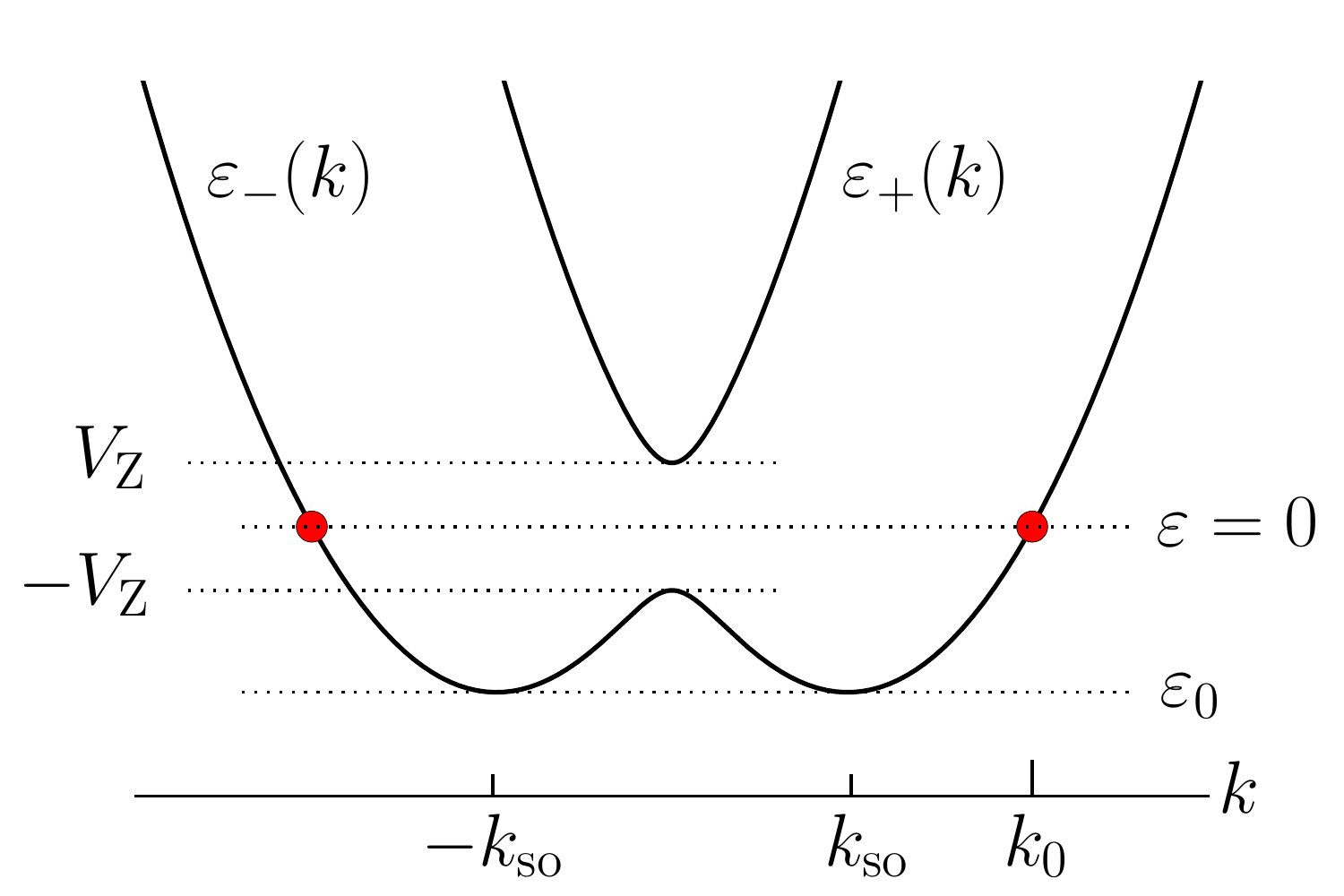}
\caption{\label{Dispersion} Dispersion relation in a one-dimensional wire in the presence of Rashba spin-orbit and Zeeman interaction. The gap at $k=0$ is entirely due to the Zeeman energy $V_{\rm Z}$. For $\alpha k_{\rm F}\gg V_{\rm Z}$ the position of the two minima  $\varepsilon=\varepsilon_0$ is approximately given by $\pm k_{\rm so}\equiv\alpha m/\hbar^2$. The topological regime requires having the chemical potential lying inside the gap, as shown here; $k_0$ denotes the point at which the dispersion crosses the representative mid-gap  level $\varepsilon=0$.}
\end{figure}
Then, assuming the most favorable situation $\alpha k_{\rm F}\gg V_{\rm Z}$ (not so easy to achieve experimentally yet~\cite{Kouwenhoven2012Science}) and thus $\Delta_{\rm T}\simeq\Delta_{\rm pr}$, the requirement of a minimum topological gap of 100~mK translates into the condition $\Gamma_0 \simeq\Delta_{\rm pr} \simeq~$100~mK. 

As a final step, we need to estimate the normal-state density of states $\nu^{\rm n}_{\rm T}(0)$ in the topological wire. To do so, we use the fact that the desired chemical potential has to lie in-between the gap opened by the Zeeman interaction added to the Rashba helical bands (at least in the simplest, ideal one-channel model). Using the dispersion relation 
\begin{align}
\varepsilon_\pm (k)={\hbar^2 k^2}/{2m}\pm \sqrt{V_{\rm Z}^2 + \alpha^2 k^2} -\mu 
\end{align}
and requiring that the chemical potential lies in the middle of the Zeeman gap, for instance halfway between $\varepsilon_-(0)$ and $\varepsilon_+(0)$ (as shown with red points in Fig.~\ref{Dispersion}), one gets the simple condition $\mu=0$. 
The 1D density of states per unit volume at this energy is
\begin{align}\label{rho-Rashba}
\nonumber \bar \nu^{\rm n}_{\rm T}(\varepsilon=0)&=\left.\frac{2}{{\rm d}\varepsilon_-(k)/{\rm d}k}\right\arrowvert_{\varepsilon=0}\\
&= \frac{2}{\left(\frac{\hbar^2 k}{m}- \frac{\alpha^2 k}{\sqrt{V_{\rm Z}^2 + \alpha^2 k^2} } \right)}_{k=k_0}\;,
\end{align}
where $k_0$ satisfies $\varepsilon_-(k_0)=0$, see Fig.~\ref{Dispersion}. Insertion of the expression for $k_0$ in Eq.~(\ref{rho-Rashba}) leads to 
\begin{align}\label{rho-Rashba-chi}
\bar\nu^{\rm n}_{\rm T}(\varepsilon=0)=
\frac{\sqrt{2}}{\alpha} \frac{ \sqrt{1+\sqrt{1+\chi^2}} }{ \sqrt{1+\chi^2} }\;,
\end{align}
with $\chi$ defined above. In the considered limit $\alpha k_{\rm F}\gg V_{\rm Z}$ and thus $\chi\ll1$ the  density of states per unit volume is approximately given by
\begin{align}\label{rho-Rashba-final}
 \bar\nu^{\rm n}_{\rm T}(\varepsilon=0)\simeq\frac{{2}}{\alpha} \;.
\end{align}
The spin-orbit interaction strength $\alpha$ ranges from 0.00075 eV$\cdot$\AA~in GaAs quantum wells~\cite{Rashba1} to 0.1 eV$\cdot$\AA~in InGaAs quantum wells~\cite{Rashba2}, or even more in heavier-element-wires such as InSb~\cite{Kouwenhoven2012Science}.
We can therefore conclude that $\bar\nu^{\rm n}_{\rm T}(0)$ in the simple one-channel case varies between 10 and $10^3$ ($\mu$m$\cdot$K$)^{-1}$. We choose the average value of $\sim 10^2 ~(\mu$m$\cdot$K$)^{-1}$ and a typical wire length of $1~ \mu$m~\cite{Kouwenhoven2012Science}.
By imposing the constrain $\Gamma_0\sim 100$ mK derived above, we obtain via Eq.~(\ref{Gamma0-RT}) the final estimate for the tunnel resistance
\begin{align}
R_{\rm T}\simeq 100~\Omega\;.
\end{align}
As calculated in the former sections this value corresponds to a quasiparticle tunnel rate of $\Gamma_{\rm qp}\sim 1-10~$MHz, {\it i.e.} poisoning times of the order of $\mu$s or less, which has to be compared with the typical time required for adiabatic qubit manipulation.
The natural time scale which identifies the adiabatic regime is provided by the inverse topological gap
\begin{align}\label{adiabaticity time}
\tau_{\rm ad}=\frac{\hbar}{\Delta_{\rm T}} \;,
\end{align}
which for the aforementioned reasonable estimate of minimum gap $\Delta_{\rm T}=100~$mK takes the value $\tau_{\rm ad}\simeq 1$ ns. The requirement of adiabatic computation is then satisfied if operations are performed on a time scale $\tau_{\rm comp}$ much longer than $\tau_{\rm ad}$. In turn, quasiparticle poisoning events must be rare events during the time of computation:
\begin{align}\label{time scales}
\tau_{\rm ad}\ll \tau_{\rm comp} \ll \tau_{\rm qp}\;,
\end{align}
where we have introduced for convenience the quasiparticle poisoning time $\tau_{\rm qp}\equiv1/\Gamma_{\rm qp}$.
Assuming an order of magnitude difference between successive time scales, the above condition Eq.~(\ref{time scales}) sets the upper limit for $\Gamma_{\rm qp}$ to 10 MHz, which is in the range of values we found in our calculations for an average situation. This shows again that the phenomenon of quasiparticle poisoning is not at all marginal, and its relevance should be assessed case by case.\\

For example, for the only experimental results available so far (Ref.~[\onlinecite{Kouwenhoven2012Science}]), the proximity effect is not very effective and the observed proximity gap is about one tenth of the bulk superconducting gap (which is however large in this case). 
On top of that, the spin-orbit energy is much smaller than the Zeeman energy in the topological phase, reducing the topological gap by an additional factor (approximately a factor 5 at the onset of the topological transition).

Note that $\tau_{\rm ad }$ is set by the value of $\Delta_{\rm T}$, whereas $\tau_{\rm qp}$ is ultimately
determined by $\Delta_{\rm pr}$ (via $\Gamma_0$ and $R_{\rm T}$) and does not depend on the physical properties of the topological nanowire (except for the density of states contained in $R_{\rm T}$). Hence, the parameter regime $\alpha k_{\rm F}\ll V_{\rm Z}$ is less favorable, not only due to the fact alone that one gets smaller values of the topological gap, but also because the adiabatic time scale is increased while the poisoning time remains constant.
\\

Working in the multi-channel regime would even be less favorable, since the  density of states $\bar\nu^{\rm n}_{\rm T}$ in the wire would be noticeably increased, and to maintain the same $\Gamma_0$ the tunnel resistance  $R_{\rm T}$ should be further decreased.
A larger value of $\alpha$ would instead help in this direction, since it lowers $\bar\nu^{\rm n}_{\rm T}$ (beyond increasing the topological gap).\\

Also in the opposite limit of a transparent interface, $\Gamma_0\gg\Delta_{\rm S}$, where the proximity gap is essentially given by $\Delta_{\rm S}$, decreasing the quasiparticle tunnel rate is difficult.
Equation (\ref{Gamma0-RT}) tells us again that for $\Gamma_0\gtrsim \Delta_{\rm S}\sim 1$ meV, in order to suppress the factor  ${R_{\rm Q}}/{R_{\rm T}}$ one would need unrealistically low values of the wire density.\\

A possible improvement could be provided by the finite charging energy of the nanowire, which raises the energy of all the states and lifts the huge degeneracy of quasiparticle states close to $\Delta_{\rm T}$. For a single pair of Majorana states, the charging energy also introduces an undesired splitting between the filled and unfilled zero-energy state. But one can then work with two wires and four Majorana states, two of which remain degenerate even in the presence of a charging energy~\cite{HalperinPRB2011}.

\section{Conclusions}
In summary, we have calculated the tunnel rate (``poisoning'') of quasiparticles from a bulk superconducting reservoir to a semiconducting nanowire, which becomes also superconducting due to proximity effect. Under appropriate conditions, the nanowire is in a topological superconducting state, hosting a Majorana state at each of its ends, which could be used for topological computation. 
Using quantitative results from recent experiments on the density of excess quasiparticle in superconductors, we have shown that the poisoning of the wire could represent a serious problem, with Majorana-qubit lifetimes which range from 10 ns to 0.1 ms, depending on many physical parameters. Since some of these parameters cannot simply be adjusted independently, finding a suitable configuration which minimizes the poisoning phenomenon requires a fine-tuning of the coupled nanowire-superconducting system more delicate than one could have expected.

\section{Acknowledgments}
We thank Luka Trifunovic for useful help with the numerical calculations. 
This work has been supported by the Swiss SNF, NCCR Nanoscience, NCCR QSIT, and the EU project SOLID.

\end{document}